# A Contrastive Learning Based Convolutional Neural Network for ERP Brain-Computer Interfaces

Yuntian Cui, Xinke Shen, Dan Zhang. Chen Yang

*Abstract*—**Objective. ERP-based EEG detection is gaining increasing attention in the field of brain-computer interfaces. However, due to the complexity of ERP signal components, their low signal-to-noise ratio, and significant inter-subject variability, cross-subject ERP signal detection has been challenging. The continuous advancement in deep learning has greatly contributed to addressing this issue. This brief proposes a contrastive learning training framework and an Inception module to extract multi-scale temporal and spatial features, representing the subject-invariant components of ERP signals. Specifically, a base encoder integrated with a linear Inception module and a nonlinear projector is used to project the raw data into latent space. By maximizing signal similarity under different targets, the inter-subject EEG signal differences in latent space are minimized. The extracted spatiotemporal features are then used for ERP target detection. The proposed algorithm achieved the best AUC performance in single-trial binary classification tasks on the P300 dataset and showed significant optimization in speller decoding tasks compared to existing algorithms.**

## I. INTRODUCTION

A Brain-Computer Interface (BCI) establishes a direct communication pathway between the brain and external devices, aiming to decode brain signals into computer-understandable commands[1]. This technology finds applications in neurorehabilitation, disease diagnosis, assistive control, and entertainment[2], significantly enhancing human-computer interaction efficiency. Due to the non-invasive and cost-effective nature of EEG, BCI systems frequently utilize EEG for development[3]. However, while EEG signals offer high temporal resolution, their spatial resolution is relatively poor[4]. Additionally, EEG signals are subject to significant

Yuntian Cui, Chen Yang are, with School of Electronic Engineering, Beijing University of Post and Telecommunications, Beijing 100083, China
Dan Zhang is with Department of Psychological and Cognitive Sciences, Tsinghua University, Beijing 100084, China
Xinke Shen is with Department of Biomedical Engineering, Southern University of Science and Technology, Shenzhen, 518055, China.
(Corresponding author:Dan Zhang, Chen Yang)

attenuation and noise interference due to the volume conductor[3] effect, resulting in a low signal-to-noise ratio (SNR) that hampers the decoding of individual brain processes.

To improve the SNR of EEG, specific tasks are often employed. Event-Related Potentials (ERPs) are phase-locked EEG changes occurring in response to specific stimuli, characterized by high SNR and distinct features[5]. ERPs enable the identification of high-level cognitive activities in the brain under specific event stimuli, proving useful in target detection[6,7], medical diagnosis[8,9], emotion computing[10], and identity authentication[11]. However, ERP components are complex, exhibit significant inter-subject variability, and require lengthy calibration times, limiting the accuracy and information transmission of ERP decoding. Therefore, a generalizable ERP component detection method for inter-subject analysis is essential[5].

Extensive research on ERP detection has focused on developing recognition algorithms[12], such as Support Vector Machines (SVM)[13] and Linear Discriminant Analysis (LDA)[14,15], which are common in ERP analysis. However, these methods face performance degradation with large datasets and specific user groups. Advanced algorithms like xDawn-based ERP feature enhancement[16,17], Riemannian geometry-based ERP classifier[16, 18], and microstate-based DCPM have been proposed, advancing BCI recognition algorithms[19].

With the advancement of deep learning, fields like pattern recognition and feature engineering have experienced rapid development, significantly impacting computer vision and natural language processing[20]. Compared to traditional algorithms, deep neural networks can automatically learn complex features from large-scale data[20]. In the EEG domain, numerous studies have integrated deep neural networks, achieving notable progress[21,22]. The classic neural network for EEG signals, known as EEGNet, employs separable convolutions to design a compact convolutional neural network for EEG feature extraction, achieving excellent results across multiple paradigms[23]. Gao et al. combined CNN and LSTM to extract temporal features, enhancing EEG classification accuracy in SSMVEP tasks[24]. Addressing the significant inter-subject variability in EEG signals, many studies have employed neural networks to learn inter-subject invariance[25].

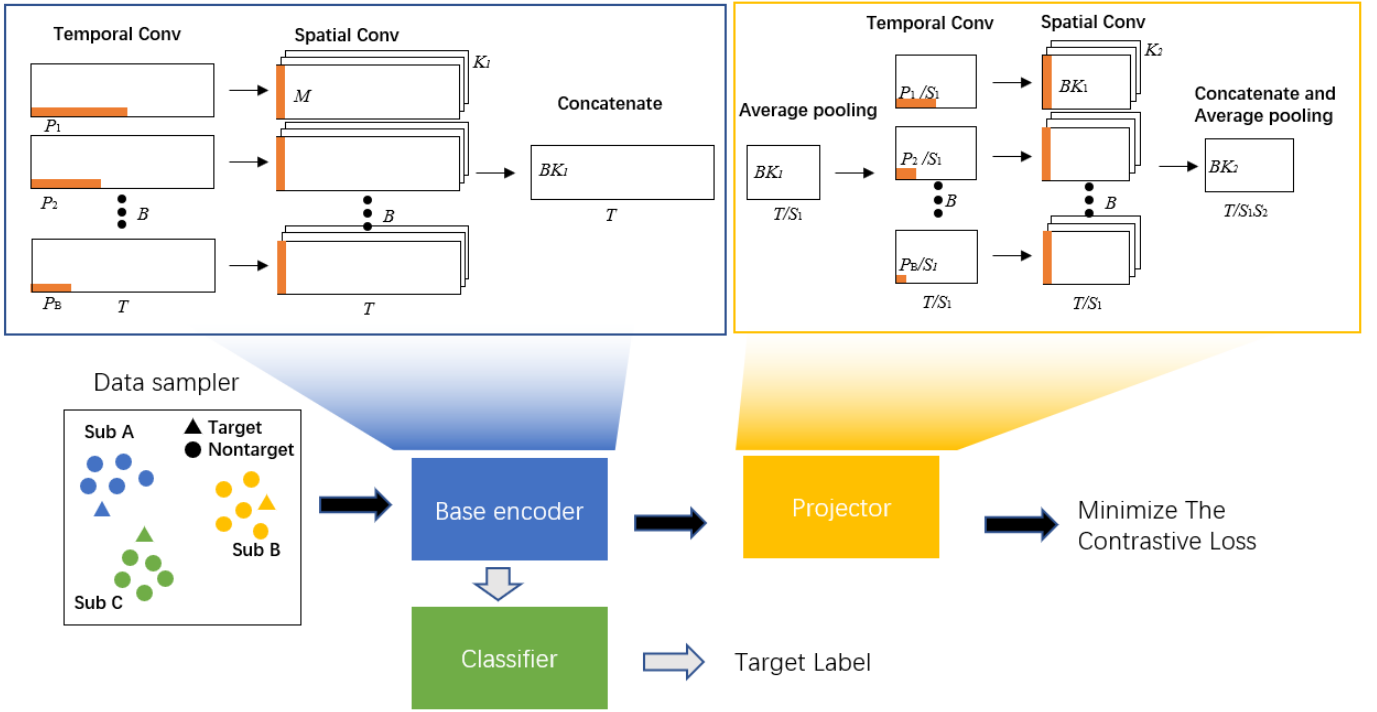

Fig. 1. The illustration of the Contrastive Learning method for Inter-subject ERP detection. In the figure, "Sub" stands for "subject" and "Conv" stands for "Convolution." Upper-left: The architecture of the base encoder . Upper-right: The architecture of the projector. The black arrows stand for the contrastive learning training procedure, and the gray arrows stand for the prediction procedure.

Sergio et al. designed a model called EEGSym, incorporating mid-sagittal symmetry, achieving favorable results in inter-subject motor imagery paradigms[26].Shen et al. developed a CNN-based contrastive learning framework that achieved state-of-the-art results in cross-subject emotion recognition, demonstrating the value of contrastive learning strategies in cross-subject scenarios[27]. Eduardo et al. implemented a network structure named EEG-Inception, utilizing the inception module from computer vision to achieve notable success in ERP recognition[28].

Despite these advancements, the variability of EEG signals across subjects necessitates the extraction of subject-invariant components for robust EEG decoding algorithms. Few studies have focused on the subject-invariant components of ERP as the primary research objective. Extracting these invariant components and generalizing them to new subjects and tasks remains an unresolved challenge. This brief proposes a model combining Inception modules and contrastive learning frameworks for cross-subject ERP component detection. The primary objectives of this brief are:
1. Proposing a contrastive learning training framework for aligning ERP features across subjects.
2. Designing a network structure optimized for maximizing inter-subject differences.
3. Exploring subject-invariant ERP components across different tasks.

## II. METHODS

The framework of our model is illustrated in Fig. 1. As mentioned in the literature[27], a contrastive learning framework typically consists of four components: a data sampler, a base encoder, a projector, and a contrastive loss function. First, the data sampler generates mini-batches composed of ERP data from different subjects. These mini-batches are used to train the subsequent base encoder to maximize inter-subject similarity. The projector maps the output of the previous module to compute their similarity, optimizing both the base encoder and projector to minimize the contrastive loss.

### A. Data Sampler

Single-trial ERP data often have low SNR. To address this, an averaged data augmentation method is employed, although the final prediction targets single trials. Therefore, the number of averaged trials should not be excessively high. To compare invariant components across different subjects, a sample pair traversal strategy similar to that in the literature[27] is adopted, traversing all subject pairs and selecting one or several averaged samples from each subject to form the mini-batch. We considered the ERP pairs within the subject pairs as positive samples.Since ERPs are evoked by the oddball paradigm, one ERP sample and five non-ERP samples are empirically chosen from each subject.

### B. Base Encoder

The base encoder takes mini-batches sampled from subject pairs as input to generate enhanced components of different subjects' ERPs. It includes an inception module, which contains three distinct branches, each with a linear temporal convolution layer and a linear spatial convolution layer. For each branch:

*Temporal Filter*: One of the key features of EEG signals is their temporal characteristics, especially the phase-locked nature of ERP components. Thus, temporal filters are applied to extract features from different frequency bands of the EEG signal. Different scales of temporal filters are used in different branches to activate features at various frequencies.The temporal filter can be formulated as

$$H_{b,k_1} = T_b * X, \ b = 1, 2, \ldots B; k_1 = 1,2, \ldots, K_1 \quad (1)$$

where $X \in \mathbb{R}^{M \times N}$ is one ERP trial, $M$ is the number of EEG channels, and $N$ is the number of time points. $B$ is the number of branchs in the Inception module. $T \in \mathbb{R}^{P_b \times 1}$ is the temporal filter , where $P_b$ stands for the filter length in the b-th branch.Sympol (*) stands for the convolution. The input $H_b$ is padded to ensure the output is still of length $N$ on the temporal dimension.

*Spatial Filter*: EEG signals can experience aliasing among multiple sources when transmitted from the source to the acquisition device, leading to strong correlations among EEG data from different channels. Therefore, spatial convolution is performed after the temporal filters in each branch to learn spatial patterns at different frequency bands, transforming EEG signals into latent space to identify relevant source spaces. The design of the spatial convolution module is inspired by the FBCSP algorithm, which assumes different spatial patterns for EEG signals across various frequency bands. The spatial filter can be formulated as

$$H_b = W_b H_{b,k_1}, b = 1, 2, \ldots B \quad (2)$$

Where $W_b \in \mathbb{R}^{M \times 1}$ is the weights of the channels in b-th branch.

The outputs of different branches are ultimately merged as

$$H = Concat(H_1, H_2, \ldots, H_b), b = 1, 2, \ldots B \quad (3)$$

Where the fused frequency features corresponding to each branches, which will be further processed by projector.

### C. Projector

A nonlinear projector is employed between the base encoder and the final contrastive loss function. The nonlinear projector better learns representations for downstream prediction tasks. Inspired by the literature[28], the projector adopts a structure similar to the base encoder, utilizing an inception module for multi-scale temporal and spatial convolution. Unlike the encoder, each convolution layer in the projector is followed by batch normalization for feature normalization mapping, an activation function to introduce nonlinearity, and regularization to prevent overfitting. Specifically:

*Average pooling:* The average pooling has a kernel size of 1 × $S_1$ and a stride of $S_1$.And the shape of the output from this layer is $BK_1 \times T/S_1$.

*Temporal Filter and Spatial Filter:* In projector, the inception module shares similar computations with base encoder.The number of brunches is also $B$. Unlike the base encoder, due to the averaging pooling layer passed before this inception module, this results in the need for the temporal filter in this inception module of the projector to vary with $S$.And after the temporal filter and spatial filter, an exponential linear unit (ELU) is used as activation function to introduce nonlinearity.These two filters reduce the parameter size and the inception module ensure specific spatio-temporal pattern extractions for each frequency band. Then the output of projector is vectorized for further similarity calculation.

### D. The Contrastive Loss

Once the input samples pass through the base encoder and projector, The similarity of the single-sample EEG trials to subject pair {A,B} in the latent space can be expressed as:

$$sim^{A,B} = \frac{h^A h^B}{\|h^A\| \|h^B\|} \quad (4)$$

Where h is one-dimensional vector as output of projector. The goal of contrastive learning is to increase the similarity between positive sample pairs within each mini-batch and decrease the similarity between negative sample pairs, which aligns with the basic principles of many traditional EEG decoding algorithms. Similar to the SimCLR framework[29], normalized temperature-scaled cross-entropy loss is employed. The cross-entropy loss between two samples is represented by().

$$l^A = -\log \frac{\exp(sim^{A,B}/\tau)}{\sum \Delta \exp(\exp(sim^{A,B}/\tau)) + \sum \exp(sim^{A,B}/\tau)} \quad (5)$$

Where $\tau$ is the temperature coefficient in contrastive learning,and $\Delta \in \{0, 1\}$ is an an indicator function. It is set to 1 if positive pairs. By minimizing the loss function, the model will increase the similarity of the positive pair in subject pair {A,B}.And the overall loss function for the entire mini-batch is calculated as:

$$L = \sum l^A + \sum l^B \quad (5)$$

### E. Classifier

In the prediction procedure, the trained base encoder is used to extract inter-subject ERP components from the data. Subsequently, a classifier is defined and trained, consisting primarily of two convolutional layers with gradually decreasing filter numbers, designed to learn more specific downstream classification tasks. Since ERP waveforms are phase-locked, it is assumed that the main features of ERP are distributed in the temporal domain. Finally, a linear fully connected layer maps the temporal EEG signals in latent space directly to the classification target.

## III. EXPERIMENTAL RESULTS

### A. Dataset

The P300-based speller dataset comprises EEG signals from 73 subjects, including 42 healthy controls and 31 motor-disabled subjects[28]. The dataset was collected using eight electrodes placed at Fz, Cz, Pz, P3, P4, PO7, PO8, and Oz in the occipital region of the brain, with a sampling rate of 256Hz, subsequently downsampled to 128Hz. In all cases, subjects utilized an ERP-based speller using the row-column paradigm. This paradigm displays a command matrix where rows and columns randomly highlight[30]. When a subject selects a command, they are required to gaze at the chosen command, and ERPs are elicited upon perceiving the stimulus. The system decodes the EEG signals using signal processing algorithms to locate the selected target and provide feedback to the subject.

The number of decoding targets in the dataset ranges from 16 to 54. For this experiment, data from 7 randomly selected subjects (including both healthy and motor-disabled subjects) were used as the test set, while the remaining subjects were used as the training set. To reduce training time, each subject in the training set selected 60 target trials and 300 nontarget trials. Detailed information about data acquisition is available in the literature[28].

### B. Parameter Settings:

#### 1) Data sampler:

Data augmentation was performed using averaged trials since single-trial ERP data contain high random noise, which could significantly impair the training efficiency of the contrastive learning process. However, excessive averaging of trials would prevent the base encoder from recognizing single-trial EEG signals. Thus, 3 trials were averaged to enhance SNR while retaining the network's ability to process single-trial signals.

#### 2) Inception module:

We use the three-layer inception module for subsequent feature extraction empirically. Temporal filter convolution kernels in the three branches of the Inception module correspond to fs/2, fs/4, and fs/8, respectively. The lengths of the convolution kernels were set to 64, 32, and 16 to capture frequency domain features above 2Hz, 4Hz, and 8Hz. Each branch's kernel number was set to 8. In the projector, the average pooling layer size was set to (1, 4), while in the classifier, it was set to (1, 2). The activation function was empirically chosen as ELU. During the contrastive learning phase, layered normalization was applied, connecting and normalizing the same channel's different samples from the same subject within a mini-batch.

The contrastive learning algorithm was implemented using the PyTorch framework. For optimizing the contrastive learning model, the model was trained for 100 epochs with early stopping (a maximum tolerance of 30 epochs without validation accuracy improvement). The Adam optimizer was used for all network parameters, with the learning rate and weight decay empirically set to $10^{-3}$ and 0.015, respectively. For optimizing the classifier model, the learning rate was set empirically at 0.0005, and the model was also trained for 100 epochs with early stopping.

The cross validation technique were used to evaluate the performance of our study in a subject-independent ERP detection task. Specifically, the data of 7 subjects were first selected as the testing set. Among the remaining 66 subjects, 5 times of 50-fold cross-validation was randomly conducted, i.e., each subject will get 25 classification results, and each algorithm will have 7*25 test set results to evaluate the algorithm performance.

### C. Results

This brief initially compares the performance of four existing algorithms in the binary classification task of ERP target detection: LDA, xDawn+RG, EEGNet, and EEGInception.

Given that the number of target samples is much smaller than the number of non-target samples, AUC (Area Under the Curve) is chosen as the evaluation metric for the binary classification. The detection results on the RSVP-based benchmark dataset are shown in Fig. 2.

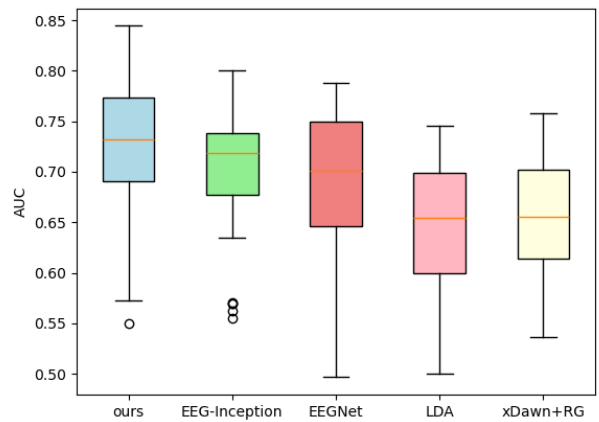

Fig. 2. Comparison results with four methods in the subject-independent ERP detection task.

Compared with the deep learning based methods, the performance of LDA is the worst. The main reason is that handcrafted features are difficult to extract the important information of EEG signals. For traditional algorithms, it is often challenging to uncover the complex subject-invariant features, leading to significantly lower performance compared to neural network-based algorithms. The AUCs of our method, EEGInception, and EEGNet are 0.7233±0.0750, 0.7013±0.0681, and 0.6807±0.0851, respectively, significantly outperforming the two traditional algorithms. This indicates that these models successfully learn the spatial and temporal representations of ERP signals. Compared to EEGInception and EEGNet, our algorithm improves the AUC by 3.04% and 5.88%, respectively, demonstrating that our method can more effectively extract key spatiotemporal features through contrastive learning. It is proved that our algorithm has a very significant improvement in the performance of existing algorithms($p < 0.01$). These experiments successfully validate the superiority of our proposed algorithm for ERP target recognition on a public dataset.

Additionally, this brief extends the analysis to decode single-trial EEG data for the P300 speller to demonstrate the optimization effects of the proposed algorithm in practical application scenarios. The accuracy results of the speller task are shown in Table I.

TABLE I
RESULTS OF DIFFERENT ALGORITHMS ON THE SPELLER TASK

| Alorgrithm | Accuracy |
|---|---|
| Our Method | 22.65±6.35 |
| EEG-Inception | 19.65±7.31 |
| EEGNet | 19.13±5.28 |
| LDA | 14.47±5.98 |
| xDawn+RG | 16.70±6.05 |

The results are generally consistent with those reported in the literature[28]. Single-trial decisions often fail to meet accuracy requirements, necessitating repeated experiments to ensure information transmission accuracy. Although there is a statistically significant improvement compared to existing algorithms, the current algorithm research is still insufficient to achieve satisfactory discrimination in single-trial cross-subject judgments. Additionally, significant performance variability among different subjects results in high algorithm variance.

## IV. Conclusion

In this brief, a contrastive learning-based training framework for cross-subject ERP detection is proposed to decode and analyze EEG signals from different subjects. Data augmentation is performed through a data sampler, and a base encoder integrated with Inception modules and a projector is used to enhance the similar components of EEG signals across different subjects. The ERP features of EEG signals are learned and projected into latent space, and a classifier is trained to classify the extracted features. The proposed algorithm is validated on a public dataset, showing its effectiveness in improving ERP target detection classification performance and speller classification accuracy, significantly reducing pre-calibration time for new users. Future work will continue to explore contrastive learning-based ERP signal decoding algorithms.